\documentclass[aps,prc,preprint,amsmath,amssymb,showpacs,preprintnumbers,superscriptaddress]{revtex4-1}
\usepackage{CJK}
\usepackage{graphicx}
\usepackage{dcolumn}
\usepackage{bm}
\usepackage{color}
\usepackage{hyperref}
\allowdisplaybreaks[4]

\begin{document}
\begin{CJK*}{UTF8}{}

\title{Sensitivity study of \emph{r}-process abundances to nuclear masses}
\author{X. F. Jiang \CJKfamily{gbsn} (姜晓飞)}
\affiliation{State Key Laboratory of Nuclear Physics and Technology, School of Physics, Peking University, Beijing 100871, China}

\author{X. H. Wu \CJKfamily{gbsn} (吴鑫辉) }
\affiliation{State Key Laboratory of Nuclear Physics and Technology, School of Physics, Peking University, Beijing 100871, China}

\author{P. W. Zhao \CJKfamily{gbsn} (赵鹏巍)}
\email{pwzhao@pku.edu.cn}
\affiliation{State Key Laboratory of Nuclear Physics and Technology, School of Physics, Peking University, Beijing 100871, China}

\begin{abstract}
The impact of nuclear mass uncertainties on the \emph{r}-process abundances has been systematically studied with the classical \emph{r}-process model by varying the mass of every individual nucleus in the range of $\pm0.1$ to $\pm3.0\ \mathrm{MeV}$ based on six different mass models. 
A new quantitative relation between the uncertainties of \emph{r}-process abundances and those of the nuclear masses is extracted, i.e., a mass uncertainty of $\pm0.5\ \mathrm{MeV}$ would lead to an abundance uncertainty of a factor around 2.5. 
It is found that this conclusion holds true for various mass models. 
\end{abstract}


\maketitle

\end{CJK*}


\section{Introduction}\label{Section1}

The rapid neutron capture process (\emph{r}-process) in an explosive astrophysical environment has been believed to be responsible for the nucleosynthesis of about half of the elements heavier than Iron for more than half a century~\citep{Burbidge1957RMP}. 
However, it remains as one of the most challenging topics due to the uncertainties from both astrophysics and nuclear physics~\citep[]{Arnould2007PR, Thielemann2011PPNP, Thielemann2017ARNP, Kajino2019PPNP, Cowan2021RMP}. 
Neutron-rich jets from core-collapse supernovae~\citep[]{Meyer1992APJ, Woosley1994APJ, Qian1996APJ, Wanajo2001APJ} and neutron star mergers~\citep[]{Symbalisty1982APL, Freiburghaus1999APJL, Wanajo2014APJL} are two attractive potential \emph{r}-process sites. 
However, an unambiguous identification of the dominant sites of the \emph{r}-process nucleosynthesis has remained elusive, and the exact astrophysical conditions are even more difficult to know. 
Aside from the astrophysical conditions, reaction properties of thousands of neutron-rich nuclei are required in \emph{r}-process simulations, and most of these nuclei lie far beyond the experimentally known region. 
Theoretical predictions of the properties of these neutron-rich nuclei have large uncertainties, which can further interplay with the astrophysical uncertainties. This makes it even more difficult to address the issues of the \emph{r}-process.

Nuclear masses are extremely important for \emph{r}-process simulations, as they are needed to extract the reaction energies that go into the calculations of all involved nuclear reaction rates.
During recent decades, great achievements in nuclear mass measurements have been made thanks to the development of radioactive ion beam facilities, and about 2500 nuclear masses have been measured to date~\citep{Wang2012CPC, Wang2017CPC, Wang2021CPC}.
Nevertheless, the masses of a large number of neutron-rich nuclei relevant to the \emph{r}-process remain unmeasured and cannot be reached in the foreseeable future, due to the difficulties in production, separation, and detection.
Therefore, reliable theoretical predictions for nuclear masses are imperative at present.
However, an accurate prediction of nuclear masses is still a great challenge for nuclear theory due to the difficulties in understanding nuclear interactions and quantum many-body systems.
Tremendous efforts have been made in improving the prediction of nuclear masses and, currently, the root-mean-square (rms) deviations between theoretical mass models~\citep{Wang2014PLB, Moller2016ADNDT, Goriely2009PRL, Pearson1996PLB, Zuker2008RFMS, Koura2005PTP} and the available experimental data reach about $0.5$ MeV. 
The rms deviations are further reduced to about $0.2$~MeV by employing machine learning approaches~\citep[]{Wang2011PRC, Niu2013PRC, Utama2016PRC, Niu2018PLB, Neufcourt2018PRC, Wu2020PRC}. 
However, different mass models predict very different masses when moving to the region of neutron-rich nuclei and, thus, lead to obviously different abundances of nuclei in the \emph{r}-process nucleosynthesis simulations~\citep[]{Wanajo2004APJ, Sun2008PRC, Arcones2011PRC, Farouqi2010APJ, Martin2016PRL, Li2019SC, Zhao2019APJ}.

Therefore, it is important to address how the nuclear mass uncertainties limit our knowledge of the \emph{r}-process~\citep[]{Sun2008PRC, Farouqi2010APJ, Arcones2011PRC, Brett2012EPJA, Xu2013PRC, Aprahamian2014AIPA, Mumpower2016PPNP, Martin2016PRL, MendozaTemis2016JPCS, Li2019SC, Sprouse2020PRC}. 
Addressing the impact of mass uncertainties on \emph{r}-process abundances is useful to deepen our understanding of the \emph{r}-process in the present stage. For instance: Can we identify the dominant \emph{r}-process site with current uncertainty~\citep[]{Martin2016PRL, Mumpower2015PRC}? Can we understand the origin of the rare-earth peak~\citep[]{Surman1997PRL, Goriely2013PRL}? To what extent can we trust the nuclear chronometers~\citep[]{Goriely2001AA, Schatz2002ApJ, Niu2009PRC}? 

Sensitivity studies were carried out to address how the nuclear mass uncertainties impact the \emph{r}-process abundances by varying every single nuclear mass in the \emph{r}-process simulations~\citep[]{Brett2012EPJA, Aprahamian2014AIPA, Mumpower2015JPG, Mumpower2015PRC, Mumpower2016PPNP}. 
Such studies can provide a link between the uncertainty of nuclear masses and the \emph{r}-process abundances, and can identify the key nuclei that have strong impact on the \emph{r}-process abundances~\citep{Mumpower2015PRC}. 
Therefore, they are important for both the future experimental and theoretical studies of nuclear masses.

Currently, sensitivity studies with the nuclear mass variations $\Delta M = \pm0.5$ and $\pm0.1\ \mathrm{MeV}$ have been carried out~\citep{Mumpower2015EPJ, Mumpower2015PRC} based on the mass models HFB17~\citep{Goriely2009PRL}, DZ33~\citep{Kirson2012NPA}, FRDM1995~\citep{Moller1995ADNDT}, and FRDM2012~\citep{Moller2016ADNDT}. 
It turned out that mass uncertainties of $\pm0.5\ \mathrm{MeV}$ have a significant impact on \emph{r}-process abundances~\citep{Mumpower2015PRC}, while mass uncertainties of $\pm0.1\ \mathrm{MeV}$ may be enough for abundance details to be clearly resolved~\citep{Mumpower2015EPJ}. However, it is known that the differences between the nuclear masses predicted by different mass models, for neutron-rich nuclei far away from the experimental known region, can be as large as several or even several tens of $\mathrm{MeV}$. Therefore, the sensitivity study with the nuclear mass deviation $\Delta M$ larger than $\pm0.5\ \mathrm{MeV}$ is necessary to address the impact of current nuclear mass uncertainties on \emph{r}-process abundances.

In this work, we aim to systematically illustrate the impact of nuclear mass uncertainties on \emph{r}-process abundances. Simulations based on various nuclear mass models are performed and the obtained abundances are compared with the Solar \emph{r}-process abundances. 
By varying every single nuclear mass within a series of different uncertainties, the impact of individual nuclear masses on the \emph{r}-process abundances are studied, and a new quantitative relation between the \emph{r}-process abundance variations and the nuclear mass uncertainties is proposed and analyzed.


\section{Classical \emph{r}-process model}\label{Section2}

A systematical study of the impact of nuclear mass uncertainties on \emph{r}-process abundances in this work requires about one million times of \emph{r}-process simulations, which are too heavy for the dynamical $r$-process model with the nuclear reaction network simulations~\citep[]{NucNet, SkyNet}. Therefore, a site-independent \emph{r}-process model, the so-called classical \emph{r}-process model, is employed in this investigation, which can be regarded as a realistic simplification of the dynamical \emph{r}-process model, and has been successfully employed in describing \emph{r}-process patterns of both the Solar System and metal-poor stars~\citep[]{Kratz1993APJ, Kratz2007APJ, Sun2008PRC, Xu2013PRC}. 
Nevertheless, it should be noted that the real neutron freeze-out after the equilibrium between neutron capture and photodisintegration reactions, as well as the fission recycling are neglected in the present classical \emph{r}-process model. In particular, nuclear fission may play an important role in the \emph{r}-process nucleosynthesis, especially in the neutron star mergers scenarios~\citep[]{Eichler2015ApJ, Sprouse2020EPJ, Cowan2021RMP}.

In the classical \emph{r}-process model, iron group seed nuclei are irradiated by high-density neutron sources with a high-temperature $T\gtrsim 1.5\ \mathrm{GK}$. The \emph{r}-process abundances are obtained by the superposition of abundances from the simulations in 16 different neutron flows with neutron densities in the range of $10^{20}$ to $10^{27.5}\ \mathrm{cm}^{-3}$. The weight $\omega$ and the irradiation time $\tau$ of each neutron flow follow exponential relations on neutron density $n_n$~\citep{Kratz1993APJ, Chen1995PLB}:
\begin{align}
\begin{split}\label{eq1}
   \tau(n_n) = a\times n_n^b, \\
   \omega(n_n) = c\times n_n^d.
\end{split}
\end{align}
The parameters $a$, $b$, $c$, and $d$ can be determined from a least-squares fit to the Solar \emph{r}-process abundances.

In the astrophysical environments with high-temperature $T\gtrsim 1.5\ \mathrm{GK}$ and high neutron density $n_n\gtrsim 10^{20}\ \mathrm{cm}^{-3}$, the equilibrium between neutron capture and photodisintegration reactions can be achieved, and the abundance ratios of neighboring isotopes on an isotopic chain can be obtained by the Saha equation~\citep[]{Cowan1991PR, Qian2003PPNP, Arnould2007PR}:
\begin{equation}\label{eq2}
   \frac{Y(Z,A+1)}{Y(Z,A)}=n_n\left(\frac{2\pi \hbar^2}{ m_\mu k T}\right)^{3/2}\frac{G(Z,A+1)}{2G(Z,A)}\left(\frac{A+1}{A}\right)^{3/2}\times \exp\left[\frac{S_n(Z,A+1)}{k T}\right],
\end{equation}
where $Y(Z,A)$, $S_n(Z,A)$, and $G(Z,A)$ are respectively the abundance, one-neutron separation energy, and partition function of nuclide $(Z,A)$, and $\hbar$, $k$, and $m_\mu$ are the Planck constant, Boltzmann constant, and atomic mass unit, respectively. Note that the neutron separation energy $S_n$ deduced from nuclear masses appears in the exponential, suggesting the importance of nuclear masses in the equilibrium.

The abundance flow from one isotopic chain to the next is governed by $\beta$ decays and can be expressed by a set of differential equations:
\begin{equation}\label{eq3}
   \frac{dY(Z)}{dt}=Y(Z-1)\sum_{A}P(Z-1,A)\lambda_\beta^{Z-1,A}-Y(Z)\sum_{A}P(Z,A)\lambda_\beta^{Z,A},
\end{equation}
where $\lambda_\beta^{Z,A}$ is the total decay rate of the nuclide $(Z,A)$ via the $\beta$ decay and the $\beta$ delayed neutron emission, and $Y(Z)=\sum_{A}Y(Z,A)=\sum_{A}P(Z,A)Y(Z)$ is the total abundance of each isotopic chain. By using equations~\eqref{eq2} and~\eqref{eq3}, the abundance of each isotope can be determined. After the neutrons freeze-out, the unstable isotopes in the neutron-rich side will proceed to the stable isotopes mainly via $\beta$ decays, and the final abundances are obtained. 


\section{Baseline \emph{r}-process simulations}\label{Section3}

The sensitivity study is based on a baseline simulation, in which the nuclear masses are taken from a given mass model, and the astrophysical trajectory, i.e., the weight $\omega$ and the irradiation time $\tau$ of neutron flows, is determined by fitting the obtained abundances to the Solar \emph{r}-process abundances~\citep{Sneden2008ARAA} at the temperature $T=1.5\ \mathrm{GK}$. 
For each simulation, the nuclear masses are taken from one of the six nuclear mass models, including WS4~\citep{Wang2014PLB}, ETFSI-Q~\citep{Pearson1996PLB}, FRDM2012~\citep{Moller2016ADNDT}, RMF~\citep{Geng2005PTP}, HFB31~\citep{Goriely2016PRC}, and DZ31~\citep{Zuker2008RFMS}, if the experimental data~\citep{Wang2012CPC} are not available. As for the $\beta$-decay rates, the predictions of the FRDM+QRPA method~\citep{Moller2003PRC} are employed together with the experimental data~\citep{Audi2003NPA}.

\begin{figure}[htbp]
  \centering
  \includegraphics[width=0.5\textwidth]{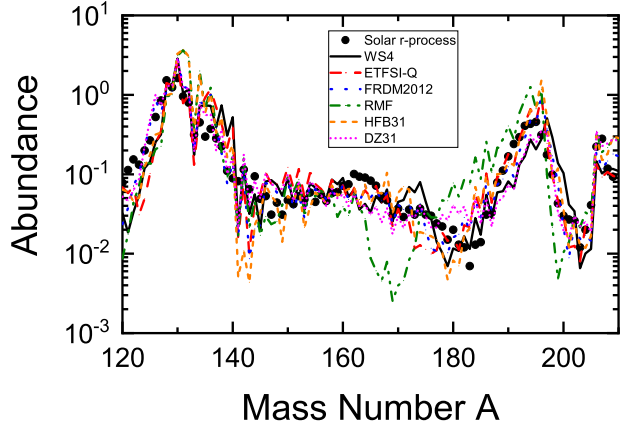}
  \caption{Baseline \emph{r}-process abundances based on different mass models as functions of the mass number $A$. The Solar \emph{r}-process abundances are displayed with black dots.}
  \label{fig1}
\end{figure}

The \emph{r}-process abundances from baseline simulations with different mass models are shown in Fig.~\ref{fig1} in comparison with the Solar \emph{r}-process abundances. Generally speaking, the overall agreement from $A=120$ to $A=209$ is satisfactory for the simulations with different mass models. 
Note that the overall agreement of the predicted abundances based on the RMF model is not as good as others, especially for the rare-earth region around $A = 170$. 
As discussed  in  Ref.~\citep{Sun2008PRC}, the deficiency is due to the fact that the neutron separation energies predicted by the RMF model increase significantly in the lanthanides region with mass number around $A = 170$ to $A=180$, which corresponds to the potentially wrongly assigned location of the shape transition before the neutron magic number in the RMF predictions. 
This deficiency may reflect the necessity to improve the relativistic density functional, e.g., the PC-PK1~\citep{Zhao2010PRC}, which provides a much better description for nuclear masses~\citep{Lu2015PRC, Yang2020CPC}.

To quantify the difference between the abundances from the baseline simulation and the Solar \emph{r}-process abundances, the abundance deviation $\Delta Y$ is introduced
\begin{equation}\label{eq4}
    \Delta Y=\sqrt{\frac{1}{N}\sum_{A=125}^{209}[\log Y^{\bigodot}(A)-\log Y^{\mathrm{pre.}}(A)]^2},
\end{equation}
where $Y^{\mathrm{pre.}}(A)$ is the abundance from the baseline simulation and $Y^{\bigodot}(A)$ is the Solar \emph{r}-process abundance. Note that $\Delta Y$ is also used in the fitting procedure that determines the astrophysical parameters in equation~\eqref{eq2}, so it stands for the smallest deviation that can be achieved in the classical \emph{r}-process model for a given set of nuclear physics inputs.

\begin{figure}[!htbp]
  \centering
  \includegraphics[width=0.5\textwidth]{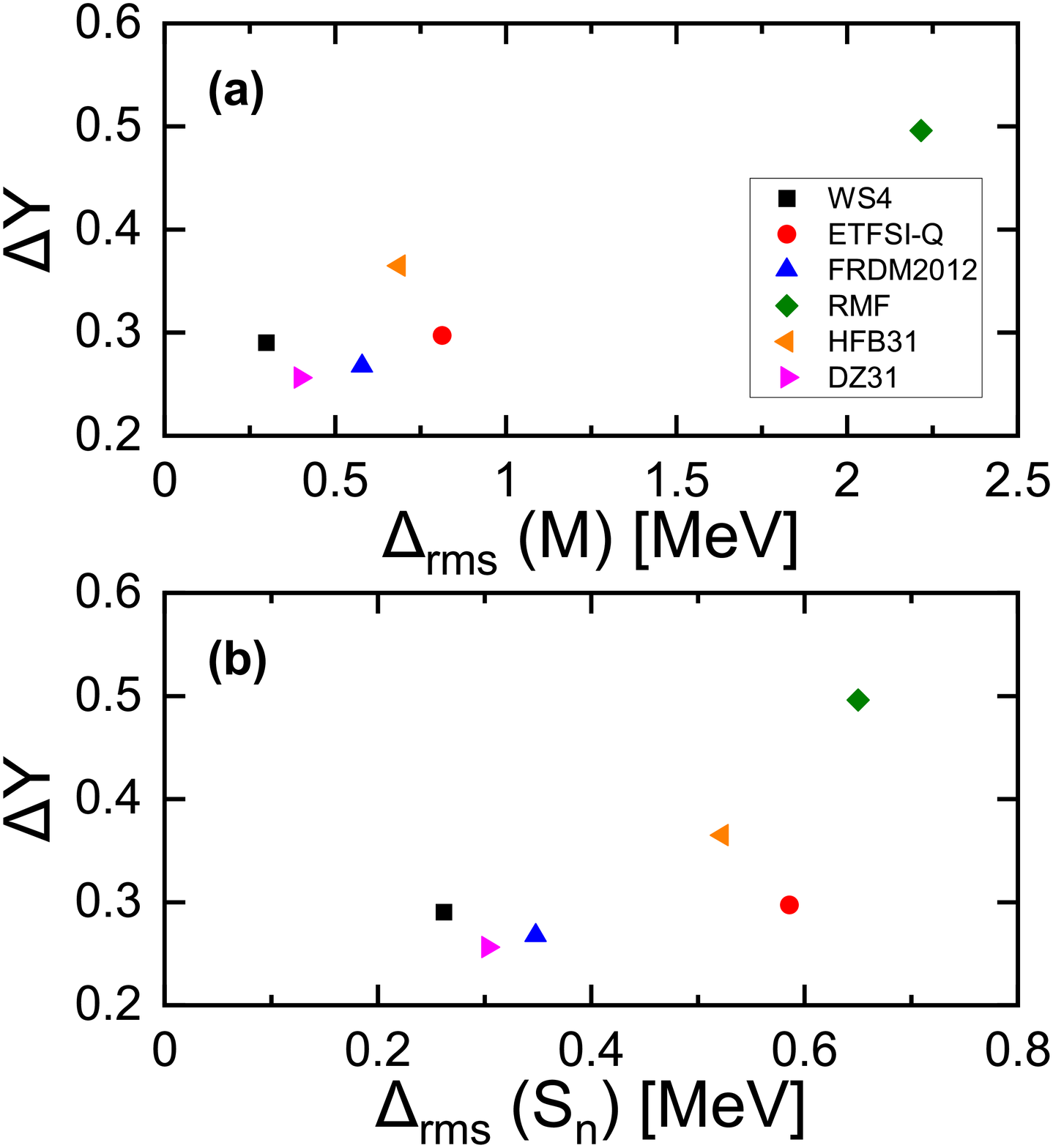}
  \caption{The abundance deviations $\Delta Y$ against (a) the rms deviations of mass $\Delta_{\mathrm{rms}}(M)$ and (b) the rms deviations of one-neutron separation energy $\Delta_{\mathrm{rms}}(S_n)$ with respect to available experimental data~\citep{Wang2012CPC} for different mass models.}
  \label{fig2}
\end{figure}
In Fig.~\ref{fig2}, it depicts the abundance deviations $\Delta Y$ against the rms deviations of mass $\Delta_{\mathrm{rms}}(M)$ and the rms deviations of one-neutron separation energy $\Delta_{\mathrm{rms}}(S_n)$ with respect to the available experimental data~\citep{Wang2012CPC} for different mass models. 
The $\Delta Y$ for the RMF mass models are obviously large due to the reasons mentioned above. 
For other mass models, although the rms deviations of mass $\Delta_{\mathrm{rms}}(M)$ range from $0.30$ MeV to $0.81$ MeV, the corresponding abundance deviations $\Delta Y$ have similar values around 0.3. 
Similar features can also be seen in Fig.~\ref{fig2}(b) against the one-neutron separation energy. 
Therefore, there seems no obvious regular relation between $\Delta Y$ and $\Delta_{\mathrm{rms}}(M)$ [or $\Delta_{\mathrm{rms}}(S_n)$]. 
This suggests that one cannot extrapolate the nuclear mass uncertainties in the experimentally unknown neutron-rich region directly from the ones in the known region.


\section{Sensitivity study of nuclear masses}\label{Section4}

\begin{figure*}[!htbp]
  \centering
  \includegraphics[width=0.9\textwidth]{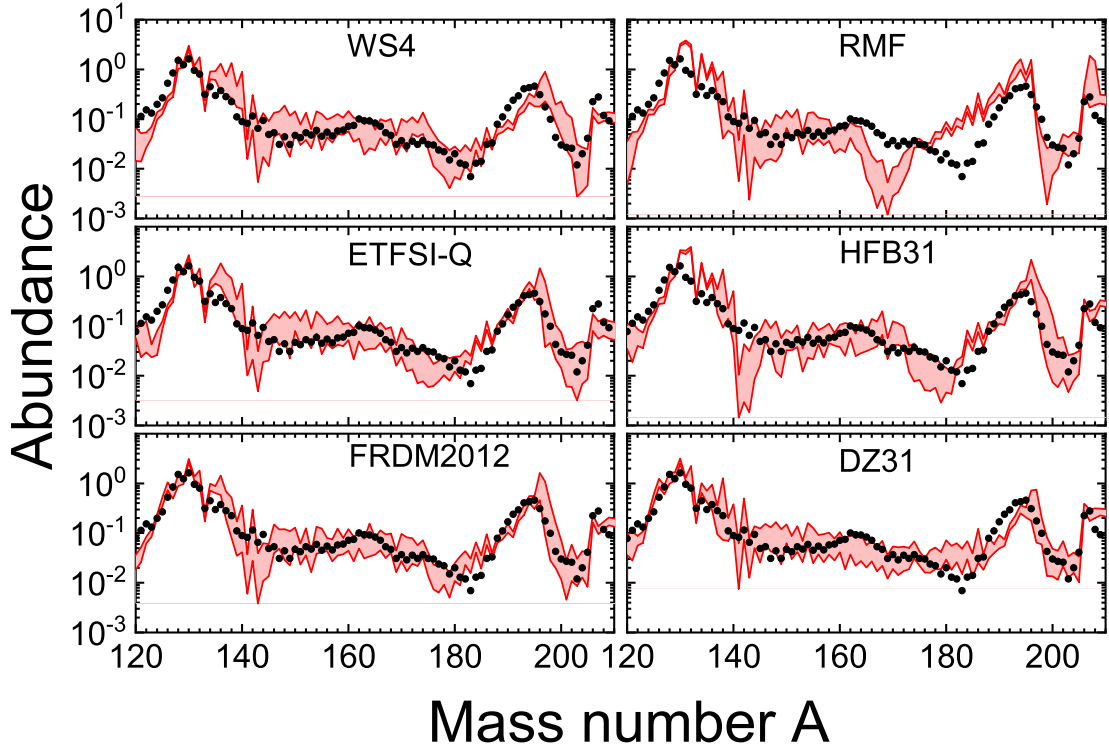}
  \caption{Variances of the \emph{r}-process abundances (shaded bands) corresponding to the sensitivity studies of $\Delta M=\pm0.5\ \mathrm{MeV}$ for various mass models.}
  \label{fig3}
\end{figure*}
Firstly, for each mass model, the sensitivity studies are carried out by varying every single predicted nuclear mass with $\Delta M =\pm 0.5\ \mathrm{MeV}$ in the \emph{r}-process simulations. The masses known from the experiment~\citep{Wang2012CPC} are not changed. 
The approximate relationship $1/T_{1/2} \propto Q^5$ is used to change the $\beta$-decay half-lives consistently with the mass variations.
Subsequently, thousands of resultant \emph{r}-process abundance patterns are obtained and they form a band as shown in each subplot of Fig.~\ref{fig3}. 
The widths of the uncertainty bands are roughly similar for different mass models. 

It is inspiring to find that some features of the final abundances are barely affected by mass uncertainties, such as the abundance peaks around $A = 130$ and $A = 195$, which are mainly due to the shell gap at $N=82$ and $N=126$, and they are much larger than $0.5$ MeV.
However, the rare-earth peak around $A = 160$, which can in principle be used to constrain the \emph{r}-process site~\citep{Mumpower2012PRC}, is entirely covered by the bands. 
This means that the uncertainty of $\pm0.5$ MeV on nuclear masses is too large to resolve this feature. 

Since the \emph{r}-process path is far away from the experimentally known region, the mass uncertainties from theoretical mass models can actually be much larger than $\pm 0.5\ \mathrm{MeV}$. 
Therefore, sensitivity studies associated with the mass uncertainties varying from  $\pm0.1$ MeV to $\pm3.0$ MeV are performed. 
In this work, we introduce the average abundance variance
\begin{equation}\label{eq5}
    W=\sqrt{\frac{1}{N}\sum_{A=125}^{209}\left[\log Y^{\mathrm{max}}(A)-\log Y^{\mathrm{min}}(A)\right]^2}
\end{equation}
to represent the abundance uncertainties quantitatively. 
Here, $Y^{\mathrm{max}}(A)$ is the abundance at the top of the uncertainty band for mass number $A$, and $Y^{\mathrm{min}}(A)$ is the one at the bottom. 

\begin{figure}[!htbp]
  \centering
  \includegraphics[width=0.5\textwidth]{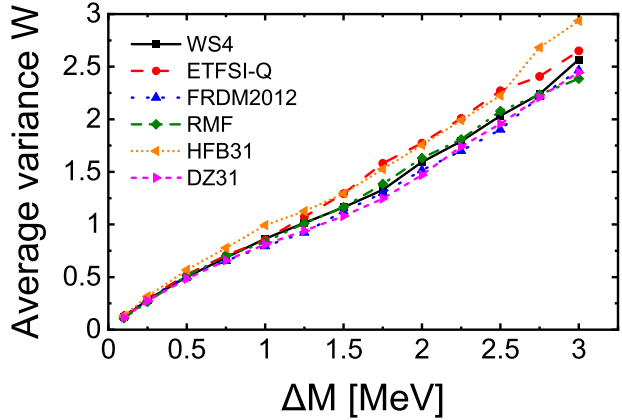}
  \caption{The average abundance variances $W$ as functions of the mass variations $\Delta M$ based on different mass models.}
  \label{fig4}
\end{figure}

In Fig.~\ref{fig4}, the average abundance variances $W$ as functions of the mass variations $\Delta M$  are shown for different mass models. 
It is found that the average abundance variances $W$ are almost linearly related to the mass variations $\Delta M$ for different mass models. 
This means that a linear change in the nuclear masses would lead to an exponential variation in the \emph{r}-process abundances. 
This feature should be related to Eq.~\eqref{eq2}, where the neutron separation energy $S_n$ deduced from nuclear masses appears in the exponential term. 
Quantitatively, if the mass uncertainties increase by $0.5\ \mathrm{MeV}$, the average abundance variance $W$ would be enlarged by about $0.38$ to $0.45$. 
Therefore, a nuclear mass uncertainty of $\pm0.5\ \mathrm{MeV}$ would lead to an abundance uncertainty of a factor about $10^{0.38}\simeq 2.40$ to $10^{0.45}\simeq 2.82$. 
Moreover, this relation is supported by the sensitivity studies based on all mass models considered in this work.


\section{Summary}\label{summary}

In summary, the impact of nuclear mass uncertainties on the \emph{r}-process abundances has been systematically studied with the classical \emph{r}-process model. 
Firstly, the baseline \emph{r}-process simulations based on six different mass models are performed. 
The obtained abundances are compared with the Solar \emph{r}-process abundances, and their differences are quantified by the abundance deviation $\Delta Y$, defined as the rms deviation of the logarithm of the abundances in the region $A=125$ to $209$. 
It turns out that there is no obvious regular relation between the abundance deviations $\Delta Y$ and the rms deviations of mass $\Delta_{\mathrm{rms}}(M)$ or the rms deviations of one-neutron separation energy $\Delta_{\mathrm{rms}}(S_n)$ for different mass models. 
However, systematic sensitivity studies showing that a nuclear mass uncertainty of $\pm0.5\ \mathrm{MeV}$ would lead to an abundance uncertainty of a factor around 2.5, and this conclusion holds true for the various mass models involved in this work.

\begin{acknowledgments}
X.F.J thanks Jie Meng for helpful discussions and guidance.
This work was partly supported by the National Key R\&D Program of China (Contracts No. 2018YFA0404400 and No. 2017YFE0116700), the National Natural Science Foundation of China (Grants No. 12070131001, No. 11875075, No. 11935003, and No. 11975031), and the High-performance Computing Platform of Peking University.
\end{acknowledgments}

\end{document}